\documentclass[showpacs,preprintnumbers,prd,showkeys,aps,twocolumn]{revtex4-1}
\usepackage{eurosym}
\usepackage{amssymb}
\usepackage{amsfonts}
\usepackage{amsmath}
\usepackage{graphicx}
\usepackage{calrsfs}
\usepackage[colorlinks=true,
            linkcolor=red,
            urlcolor=blue,
            citecolor=blue]{hyperref}

\begin{document}

\title{Quantum Effects on the Deflection of Light and the Gauss--Bonnet Theorem}
\author{Kimet Jusufi}
\email{kimet.jusufi@unite.edu.mk}
\affiliation{Physics Department, State University of Tetovo, Ilinden Street nn, 1200, Tetovo, 
Macedonia}
\affiliation{Institute of Physics, Faculty of Natural Sciences and Mathematics, Ss. Cyril and Methodius University, Arhimedova 3, 1000 Skopje, Macedonia}
\date{\today }

\begin{abstract}
In this letter we apply the Gauss--Bonnet theorem (GB) to calculate the deflection angle by a quantum corrected Schwarzschild black hole in the weak limit approximation. In particular we calculate the light deflection by two types of quantum corrected black holes: the renormalization group improved Schwarzschild solution and the quantum corrected Schwarzschild solution in Bohmian quantum mechanics. We start from the corresponding optical metrics to use then the GB theorem and calculate the Gaussian curvature in both cases. We calculate the leading terms of the deflection angle and show that quantum corrections modifies the deflection angle in both solutions. Finally by performing geodesics calculations we show that GB method gives exact results in leading order terms.
\end{abstract}
\maketitle

\section{Introduction}

According to the General Theory of Relativity, massive objects bend spacetime, as a result light rays are deflected due to the spacetime curvature. Even though the deflection of light by a massive body has a long history of research it continues even today to attract a lot of interest by studying the strong deflection regime and weak deflection regime \cite{weak1,weak2,strong1,strong2,strong3}. In this spirit, Virbhadra studied the relativistic images of Schwarzschild black hole lensing \cite{virbhadra1}, Schwarzschild black hole lensing \cite{virbhadra2}, the naked singularities and relativistic images of Schwarzschild black hole lensing \cite{virbhadra2}, gravitational lensing by traversable Lorentzian wormholes was investigated in Ref. \cite{nandi}, the light deflection by wormholes in the galactic halo region in Ref. \cite{peter}, to test the cosmic censorship hypothesis \cite{deandrea}, light deflection by charged black holes  \cite{sereno1}, the
optical effects and gravitational lensing of Kerr black holes were analyzed in Ref. \cite{bozza1,bozza2,sereno2}, gravitational lensing in braneworld gravity \cite{pal,eiroa1}, gravitational lensing by charged black holes in scalar--tensor gravity \cite{eiroa2},  Schwarzschild black hole pierced by a cosmic string \cite{cos1}, Kerr black hole pierced by a cosmic string \cite{cos2}.

Gibbons and Werner \cite{gibbons1} showed that the deflection angle can be calculated by applying the Gauss-Bonnet theorem to the optical geometry. Furthermore, this method was applied to the Schwarzschild black hole \cite{gibbons1,gibbons2}, Kerr black hole \cite{werner}, charged black hole with topological defects \cite{kimet1}, spinning cosmic strings \cite{kimet2}, and more recently the deflection angle for a finite distance for Schwarzschild--de Sitter and Weyl conformal gravity \cite{asahi}.

Modifications of the gravitational deflection of light in alternative gravity theories were considered in Ref. \cite{sumanta1,sumanta2}, strong lensing by an electrically charged black hole in Eddington-inspired Born-Infeld gravity \cite{Born-Infeld1,Born-Infeld2}. Interestig modifications of the motion of light rays were recently reported in some non-linear theories of matter, such as non-linear electrodynamics, where photons propagate over an effective geometry constructed on top of a background electromagnetic one \cite{novello}. 

However, in all above examples the light deflection was studied in the context of a classical general relativity. Thus, it's interesting to ask if one can calculate the deflection of light under the quantum effects. Interestingly enough, the answer to this question seem to be yes. In Ref. \cite{bohr}, the authors studied the bending of light in quantum gravity and shown that light deflection is modified by the quantum effects. Moreover, they investigated the possible implications of these effects of the equivalence principle. Furthermore in Refs. \cite{li,sahu} the authors have investigated the possible quantum gravity effects on the gravitational deflection of light. Motivated by the above papers, in this paper, we aim to calculate the quantum gravity effects on the light deflection via Gibbons--Werner method. To do so, we consider two quantum corrected metrics: the renormalization group improved Schwarzschild solution coming from the asymptotic safety approach that incorporates quantum corrections to the classical solution found by Bonanno and Reuter \cite{qcsh0,qcsh1,qcsh2,qcsh3} and a quantum corrected Schwarzschild metric in Bohmian quantum mechanics found by Ali and Khalil \cite{bm2} by replacing the classical geodesics with the so--called quantal or Bohmian trajectories \cite{bm1}.

This paper is organized as follows. In Section II, we write the optical metric to the improved Schwarzschild black hole and derive then the quantum corrected Gaussian optical curvature for this metric. In Section III, we apply the Gauss--Bonnet theorem and calculate the leading terms of the quantum corrected deflection angle. In Section IV, we calculate the Gaussian curvature and the deflection angle by quantum corrected Schwarzschild metric in Bohmian quantum mechanics. In Section V we confirm our results found by GW method using geodesics approach. In Section VI, we comment on our results.

\section{Renormalization group improved Schwarzschild metric}

Recently the effective improved solution coming from the asymptotic safety approach that incorporates quantum corrections was investigated by Bonanno and Reuter in Ref. \cite{qcsh1}. The key idea behind this approach is to use a running Newton constant $G=G(k)$ and to consider the standard Einstein--Hilbert action as an average action $\Gamma_{k_{obs}}$. Furthermore the scale parameter $k$ is position-dependent scale i.e. $k=k(r)$, and $r$ is the radial Schwarzschild coordinate. On the other hand, $k_{obs}$ gives a typical observational scale such that $G_0=G(k_{obs})$. The metric was found to be  \cite{qcsh1}
\begin{eqnarray}\label{1}\notag
\mathrm{d}s^{2}&=&-f(r)\mathrm{d}t^{2}+f(r)^{-1}\mathrm{d}r^{2}+r^{2}\left(\mathrm{d}\theta^{2}+\sin^{2}\theta \,\mathrm{d}\varphi^{2}\right)
\end{eqnarray}
where 
\begin{equation}
f(r)=1-\frac{2G(r)M}{r},
\end{equation}
and
\begin{equation}
G(r)=\frac{G_{0}r^{3}}{r^{3}+\tilde{\omega} G_{0}\left(r+\gamma G_{0}M\right)}.
\end{equation}

In which $G_{0}$ is Newton's gravitational constant and $M$ is the black hole mass. On the other hand $\tilde{\omega}$ and $\gamma$ are constants coming from the non--perturbative renormalization group theory \cite{qcsh0,qcsh1,qcsh2,qcsh3}. The quantum effects are encoded via $\tilde{\omega}=(167 \,\hbar) /(30 \pi)$, which shall be used later on. 
It is interesting to see that the renormalization group improved spacetime for large distances, i.e. $r\to \infty$, at order $1/r$, one can recover the Schwarzschild spacetime
\begin{equation}
f(r) = 1-\frac{2G_0 M}{r}\left(1-\tilde{\omega} \frac{G_0}{r^2}  \right),
\end{equation}
hence, the event horizon is always smaller than the Schwarzschild radius $r_{+}$. Morover this quantum corrected spacetime metric has also an inner horizon $r_{-}$, but when $M$ equals to some critical mass $M=M_{cr}$ (where $M_{cr}$, is of the order of the Planck mass) the two
horizons coincide, $r_{cr}=r_{+}=r_{-}$. While for $M<M_{cr}$, this spacetime has no horizon at all. Meanwhile the fate of singularity at $r=0$ is not clear in this approach and needs more elaboration. 

Let us briefly mention here that, Falkenberg and Odintsov, argued that the structure of effective average action depends on gauge, which suggest that the black hole solution also depends on gauge choice, this on the other hand leads to a number of problems \cite{odintsov}. To study the light deflection, let us consider the equatorial plane by choosing $\theta=\pi/2$, and then solve the metric \eqref{1} for null geodesics with $\mathrm{d}s^{2}=0$. The optical line element reads
\begin{eqnarray}\label{3}
\mathrm{d}t^{2}=\frac{\mathrm{d}r^{2}}{\left(1-\frac{2G(r)M}{r}\right)^{2}}+\frac{r^{2}\,\mathrm{d}\varphi^{2}}{1-\frac{2G(r)M}{r}}.
\end{eqnarray}

If we now introduce a new coordinate $r^{\star}$ and a new function $f(r^{\star})$,  we obtain the following form for the optical metric $\tilde{g}_{ab}$ given as \cite{gibbons1}
\begin{equation}\label{4}
\mathrm{d}t^{2}= \tilde{g}_{ab}\,\mathrm{d}x^{a}\mathrm{d}x^{b}=\mathrm{d} {r^{\star}}^{2}+f^{2}(r^{\star})\mathrm{d}\varphi^{2}; \,\,\,a,b=\lbrace r,\varphi \rbrace
\end{equation}
in which
\begin{equation}\label{5}
\mathrm{d}r^{\star}=\frac{\mathrm{d}r}{1-\frac{2G(r)M}{r}},
\end{equation}
and 
\begin{equation}\label{6}
f(r^{\star})=\frac{r}{\sqrt{1-\frac{2G(r)M}{r}}}.
\end{equation}

One can now easily check that the two nonvanishing Christoffel symbols corresponding to the optical metric \eqref{4} can be calculated as
\begin{widetext}
\begin{eqnarray}\notag
 {\tilde{\Gamma}}^{r}_{\varphi\varphi}&=& -f(r^{\star})f'(r^{\star})\\
 &=& -\frac{\left[r^6-3G_0Mr^5+2G_0r^4 \tilde{\omega}+G_0^2 M\tilde{\omega}r^3 (2\gamma-1)+G_0^2 r^2 \tilde{\omega}^2+2G_0^3 Mr\gamma+G_0^4 M^2 \gamma^2 \tilde{\omega}^2 \right]r}{\left(2G_0 r^2 M-G_0^2 M \gamma \tilde{\omega} -G_0 r \tilde{\omega}-r^3 \right)^2}
 \end{eqnarray}
 and
 \begin{eqnarray}\notag
 {\tilde{\Gamma}}^{\varphi}_{r\varphi}&=& \frac{f'(r^{\star})}{f(r^{\star})}\\
 &=& -\frac{\left[r^6-3G_0Mr^5+2G_0r^4 \tilde{\omega}+G_0^2 M\tilde{\omega}r^3 (2\gamma-1)+G_0^2 r^2 \tilde{\omega}^2+2G_0^3 Mr\gamma+G_0^4 M^2 \gamma^2 \tilde{\omega}^2 \right]}{r\left(2G_0 r^2 M-G_0^2 M \gamma \tilde{\omega} -G_0 r \tilde{\omega}-r^3 \right)(G_0^2M\gamma \tilde{\omega}+G_0 r \tilde{\omega}+r^3) }.
\end{eqnarray}
 \end{widetext}
Next to calculate the corresponding Gaussian optical curvature $K$, we make use of the only nonvanishing component of the Riemann tensor $R_{r \varphi r \varphi}$ related to the Gaussian optical curvature by the following equation \cite{gibbons1}
\begin{equation}
R_{r \varphi r \varphi}=K\left(\tilde{g}_{r \varphi}\tilde{g}_{\varphi r}-\tilde{g}_{rr}\tilde{g}_{\varphi \varphi}\right)=-K\det \tilde{g}_{r\varphi}.
\end{equation}

From this result one can show that the Gaussian optical curvature can be calculated as \cite{gibbons1}
\begin{equation}
K=-\frac{R_{r\varphi r\varphi}}{\det \tilde{g}_{r \varphi}}=-\frac{1}{f(r^{\star})}\frac{\mathrm{d}^{2}f(r^{\star})}{\mathrm{d}{r^{\star}}^{2}}.
\end{equation}

It follows now from the last equation that the Gaussian optical curvature $K$, can be expressed as \cite{gibbons1}
\begin{eqnarray}\label{11}
K&=&-\frac{1}{f(r^{\star})}\frac{\mathrm{d}^{2}f(r^{\star})}{\mathrm{d}{r^{\star}}^{2}}\\\notag
&=&-\frac{1}{f(r^{\star})}\left[\frac{\mathrm{d}r}{\mathrm{d}r^{\star}}\frac{\mathrm{d}}{\mathrm{d}r}\left(\frac{\mathrm{d}r}{\mathrm{d}r^{\star}}\right)\frac{\mathrm{d}f}{\mathrm{d}r}+\left(\frac{\mathrm{d}r}{\mathrm{d}r^{\star}}\right)^{2}\frac{\mathrm{d}^{2}f}{\mathrm{d}r^{2}}\right].
\end{eqnarray}

Using Eq. \eqref{11} and Eqs. \eqref{6} and \eqref{5} we derive the following result for the quantum corrected Gaussian optical curvature 
\begin{widetext}
\begin{equation}
K=-\frac{MG_{0}\left\lbrace 2r^9-G_0(3Mr^8+4r^7 \tilde{\omega})-\Xi(r)+\Theta(r)-r^3 \tilde{\omega}^2 M^2\gamma G_0^4(12\gamma-4)+2rG_0^5 M^2 \gamma^2 \tilde{\omega}^3 +2G_0^6 M^3 \gamma^3 \tilde{\omega}^3 \right\rbrace}{\left(r^3+ \tilde{\omega} M\gamma G_0^2+ \tilde{\omega}r G_0\right)^4},
\end{equation}
\end{widetext}
where
\begin{eqnarray}\notag
\Theta(r)&=&r^4 \tilde{\omega} MG_0^3[24 Mr\gamma-\tilde{\omega}(20 \gamma-1)],\\
\Xi(r)&=&r^5 \tilde{\omega}G_0^2[M(12 \gamma -10)r+6\tilde{\omega}].
\end{eqnarray}

Morover we can approximate the last result for the quantum corrected Gaussian optical curvature if we neglect higher order terms in $M$ to find, 
\begin{equation}
K=-\frac{2G_{0}M}{r^{3}}+\frac{12\,MG_{0}^{2}\,\tilde{\omega}}{r^{5}}+\frac{10\, M G_{0}^{3}\,\tilde{\omega}^2}{r^{7}}-\frac{24\,MG_{0}^{4}\,\tilde{\omega}^{3}}{r^{9}}.\label{14}
\end{equation}

Thus, as expected, the last equation shows that the Gaussian optical curvature is negative. This result suggests that the light deflection by a black hole should be considered as a global effect. More specifically, this brings the role of the spacetime topology on the light deflection by a black hole. In other words the  light rays can be focused only if we consider the light deflection by a black hole as a global topological effect. In the next section we are going to use the Gauss--Bonnet theorem to compute the deflection angle.

\section{Quantum corrected Deflection angle}

From differential geometry we know that the Gauss-Bonnet theorem for the non--singular region $D_ {R} $ in $M$, with boundary $\partial D_{R}=\gamma_{\tilde{g}}\cup C_ {R} $ can be stated as follows \cite{gibbons1}
\begin{equation}\label{15}
\iint\limits_{D_{R}}K\,\mathrm{d}S+\oint\limits_{\partial D_{R}}\kappa\,\mathrm{d}t+\sum_{i}\theta_{i}=2\pi\chi(D_{R}).
\end{equation}

In the last equation, $K$, is the Gaussian curvature, $\kappa$ is the geodesic curvature, $\theta_{i}$ gives the corresponding exterior angle at the $i^{th}$ vertex and $\chi(D_{R})$ is the Euler characteristic number.  Note that the geodesic curvature can be calculated as $ \kappa=\tilde{g}\,(\nabla_{\dot{\gamma}}\dot{\gamma}, \ddot{\gamma})$, in which $\tilde{g}(\dot{\gamma}, \dot{\gamma})=1$ where $\ddot{\gamma}$ is the unit acceleration vector. 

At very large $R$, i.e. $R\to \infty$, both jump angles become $\pi/2$ and hence  $\theta_{O}+\theta_{S}\to \pi$; in which $ S $ and $O$ means the source and observer, respectively (see for example \cite{gibbons1}). Furthermore, since $\gamma_{\tilde{g}}$ is a geodesic then it follows $\kappa(\gamma_{\tilde{g}})=0$. Let us find the geodesic curvature which can be calculated as $\kappa(C_{R})=|\nabla_{\dot{C}_{R}}\dot{C}_{R}|$, in which we can choose $C_{R}:= r(\varphi)=R=const$. The radial component of the geodesic curvature can be calculated as
\begin{equation}
\left(\nabla_{\dot{C}_{R}}\dot{C}_{R}\right)^{r}=\dot{C}_{R}^{\varphi}\,\left(\partial_{\varphi}\dot{C}_{R}^{r}\right)+\tilde{\Gamma}^{r}_{\varphi \varphi}\left(\dot{C}_{R}^{\varphi}\right)^{2}.
\end{equation}

Note that the first term in the last equation is zero while in the second term we have $\tilde{\Gamma}^{r}_{\varphi\varphi}=-f(r^{\star})f'(r^{\star})$ and  $\dot{C}_{R}^{\varphi}=1/f^{2}(r^{\star})$ (this result follows from $\tilde{g}_{\varphi \varphi}\,\dot{C}_{R}^{\varphi}\dot{C}_{R}^{\varphi}=1$). Finally the geodesic curvature at very large $R$, gives
\begin{widetext}
\begin{eqnarray}\notag
\lim_{R\to \infty}\kappa(C_{R})&=&\lim_{R\to \infty}\left|\nabla_{\dot{C}_{R}}\dot{C}_{R}\right|\\\notag
&=&\lim_{R\to \infty}\left\lbrace\frac{\left[R^6+G_0 \left(2R^4\tilde{\omega}-3MR^5\right)+R^2 \tilde{\omega}^2 G_0^2 \left[M(2\gamma-1)R+\tilde{\omega}\right]+G_0^4M^2\gamma^2 \tilde{\omega}^2+2G_0^3 M \gamma R \tilde{\omega}^2\right]^2}{R^2\left(R^3+G_0^2M\gamma \tilde{\omega}+G_0 R \tilde{\omega}\right)^2\left(2G_0R^2 M-G_0R \tilde{\omega}-G_0^2M\gamma \tilde{\omega}-R^3\right)^2}\right\rbrace^{1/2}\\
&\to & \frac{1}{R}.
\end{eqnarray}
\end{widetext}
As expected, we have shown that for very large $r(\varphi)=R=const.$, the geodesic curvature can be given as $\kappa(C_{R})\to R^ {-1} $.  On the other hand one can see that from the optical metric it follows  $\mathrm{d}t=R \,\mathrm{d}\,\varphi$, implying $\kappa(C_{R})\mathrm{d}t=\mathrm{d}\,\varphi$. Going back to the Gauss--Bonnet theorem \eqref{15}  and keeping in mind that the Euler characteristic is $\chi(D_{R})=1$, we find
\begin{eqnarray}
\iint\limits_{D_{R}}K\,\mathrm{d}S&+&\oint\limits_{C_{R}}\kappa\,\mathrm{d}t\overset{{R\to \infty}}{=}\iint\limits_{S_{\infty}}K\,\mathrm{d}S+ \int\limits_{0}^{\pi+\hat{\alpha}}\mathrm{d}\varphi=\pi.
\end{eqnarray}

Since we study the weak limit approximation we may choose for the light ray at zeroth order $r(t)=b/\sin\varphi $. For the deflection angle the last equation reduces to
\begin{equation}\label{19}
\hat{\alpha}=-\int\limits_{0}^{\pi}\int\limits_{\frac{b}{\sin \varphi}}^{\infty}K\,\sqrt{\det \bar{g}}\,\mathrm{d}r\,\mathrm{d}\varphi,
\end{equation}
where $\mathrm{d}S=\sqrt{\det \tilde{g}}\,\mathrm{d}r\,\mathrm{d}\varphi$, and  $\sqrt{\det\tilde{g}}=r$. Substituting the result found for the quantum corrected Gaussian curvature \eqref{14} into the last equation we end up with the following integral
\begin{widetext}
\begin{equation}
\hat{\alpha}\simeq  - \int\limits_{0}^{\pi}\int\limits_{\frac{b}{\sin \varphi}}^{\infty}\left(-\frac{2G_{0}M}{r^{2}}+\frac{10\, M G_{0}^{3}\,\tilde{\omega}^2}{r^{6}}+\frac{12\,MG_{0}^{2}\,\tilde{\omega}}{r^{4}}-\frac{24\,MG_{0}^{4}\,\tilde{\omega}^{3}}{r^{8}}\right)\mathrm{d}r\mathrm{d}\varphi
\end{equation}
\end{widetext}
Te above integral can easily be evaluated to give 
\begin{equation}
\hat{\alpha}\simeq \frac{4G_{0}M}{b}-\frac{16 M G_{0}^{2}\tilde{\omega}}{3\, b^{3}}+\frac{32 M G_{0}^{3}\tilde{\omega}^2}{15\,b^{5}}+\frac{768 M G_{0}^{4}\tilde{\omega}^{3}}{245\, b^{7}}.
\end{equation}

In the last equation $b$ is the impact parameter, therefore $b >>M$. We can see the quantum effects on the deflection angle more clearly if we use the value for the constant $\tilde{\omega}$ coming from the standard perturbative quantization of Einstein's gravity in which is shown $\tilde{\omega}$ to be \cite{qcsh0,qcsh1,qcsh2,qcsh3}
\begin{equation}
\tilde{\omega}=\frac{167 \,\hbar}{30 \pi}.
\end{equation}

Note that in the last equation we have temporary introduced the Planck constant. If we neglect the higher order terms in mass, $M$, and $\hbar$, we find
\begin{equation}
\hat{\alpha} \simeq \frac{4G_{0}M}{b}- \frac{1336 MG_{0}^{2}\hbar}{45 \,\pi \, b^{3}}+\mathcal{O}(M^{2},\hbar^{2}).
\end{equation}

The first term gives the standard deflection angle of the classical Schwarzschild black hole, whereas the second term arises due to the quantum gravity effects. Note that the factor $ MG_{0}^{2}\hbar/ (\pi b^{3}) $ present in the second term is in complete agreement with the result found in Ref. \cite{bohr}. Of course, this is a very small correction, but nevertheless modifies the deflection angle.

\section{Quantum modified Schwarzschild metric }

In a recent study \cite{bm1},  the quantum Raychaudhuri equation (QRE) in the context of the Bohmian quantum mechanics has been derived by Das. The basic idea is to replace the classical velocity field with the quantum velocity field, in particular one can introduce a four quantum velocity field $u_{\mu}$,  with the following  wave function for the quantum fluid \cite{bm2,bm2}
\begin{equation}
\psi(x^{\mu})= \mathcal{R} e^{i\,S(x^{\mu})},\label{1}
\end{equation}
where $\psi(x^{\mu})$ is a normalizable wave function, while $\mathcal{R}(x^{\alpha})$ and $S(x^{\alpha})$ are real functions. 
The four velocity field can be written as  $u_{\mu}=\frac{\hbar}{m} \, \partial_{\mu}S$ where $ (\mu=0,1,2,3)$. Later on, in Ref. \cite{bm3}  Ali and Khalil found a quantum corrected Schwarzschild metric given by (in the units  $G_{0}=c=1$)
\begin{eqnarray}\notag
\mathrm{d}s^{2}&=&-\left(1-\frac{2M}{r}+\frac{\hbar \eta}{r^{2}}\right)\mathrm{d}t^{2}+\left(1-\frac{2M}{r}+\frac{\hbar \eta}{r^{2}}\right)^{-1}\mathrm{d}r^{2}\\
&+&r^{2}\left(\mathrm{d}\theta^{2}+\sin^{2}\theta \,\mathrm{d}\varphi^{2}\right).
\end{eqnarray}

We note that Einstein's field equations are assumed to remain unchanged i.e. $G_{\mu\nu}=8 \pi T_{\mu\nu}$ with non-zero stress--energy tensor components.  If we now consider the equatorial plane $\theta=\pi/2$ and solve the above metric for null geodesics $\mathrm{d}s^{2}=0$, we find this time the optical line element to be
\begin{eqnarray}\notag
\mathrm{d}t^{2}&=&\frac{\mathrm{d}r^{2}}{\left(1-\frac{2M}{r}+\frac{\hbar \eta}{r^{2}}\right)^{2}}+\frac{r^{2}\,\mathrm{d}\varphi^{2}}{1-\frac{2M}{r}+\frac{\hbar \eta}{r^{2}}}\\
&=&\mathrm{d} {r^{\star}}^{2}+f^{2}(r^{\star})\mathrm{d}\varphi^{2},
\label{optical}
\end{eqnarray}
where 
\begin{equation}
\mathrm{d}r^{\star}=\frac{\mathrm{d}r}{1-\frac{2M}{r}+\frac{\hbar \eta}{r^{2}}},
\end{equation}
and
\begin{equation}
f(r)=\frac{r}{\sqrt{1-\frac{2M}{r}+\frac{\hbar \eta}{r^{2}}}}.\label{f}
\end{equation}

Following the same arguments as in the last section we can calculate the Gaussian optical curvature by using Eq. \eqref{11} and the last two equations to find
\begin{equation}
K=-\frac{2M}{r^{3}}\left(1-\frac{3M}{2r}\right)+\frac{3 \eta \hbar }{r^{4}}\left(1+\frac{2 \hbar \eta }{3 r^{2}}\right)-\frac{6 \hbar \eta M}{r^{5}}.
\label{curvature}
\end{equation}

Neglecting higher order terms in $M$ and $\hbar$, the above result can now be substituted into the Eq. \eqref{19} to give the following integral  
\begin{equation}
\hat{\alpha}\approx -\int\limits_{0}^{\pi}\int\limits_{\frac{b}{\sin \varphi}}^{\infty}\left(-\frac{2M}{r^{2}}+\frac{3 \eta \hbar }{r^{3}}-\frac{6 \hbar \eta M}{r^{4}}\right)\,\mathrm{d}r\,\mathrm{d}\varphi .
\label{angle}
\end{equation}

Solving this integral finally gives
\begin{equation}
\hat{\alpha} = \frac{4M}{b}-\frac{3 \pi \hbar \eta}{4 \,b^{2}}+\mathcal{O}(M^{2},\hbar^{2}).
\end{equation}

We see that the second term incorporates the quantum effects and modifies the standard deflection angle. Note that this result is in perfect agreement with the result found in Ref. \cite{bm3} where the authors have calculated the deflection angle using different method. Also one can see the similarity of this result with the deflection angle by a charged black hole \cite{kimet1}, if we set $\eta \hbar \to Q^{2}$. 

\section{Geodesics equations}
\subsection{Quantum improved Schwarzschild black hole}
Let us now confirm our results obtained by Gauss-Bonnet method by performing a standard geodesic computation using the variational principle $\delta \int \mathcal{L} \,\mathrm{d}s=0$. Using our metric (1) it follows that the Lagrangian gives 
\begin{widetext}
\begin{equation}\label{43}
\mathcal{L}=-\frac{1}{2}\left(1-\frac{2MG(r(s))}{r(s)}\right)\dot{t}^2+\frac{\dot{r}^2}{2\left(1-\frac{2MG(r(s))}{r(s)}\right)}+\frac{1}{2}r(s)^2 \left(\dot{\theta}^2+ \sin^{2}\theta \dot{\varphi}^2\right),
\end{equation}
\end{widetext}
where 
\begin{equation}
G(r(s))=\frac{G_{0}r^{3}(s)}{r^{3}(s)+\tilde{\omega} G_{0}\left(r(s)+\gamma G_{0}M\right)}.
\end{equation}

We shall consider the the deflection of light in the equatorial plane therefore we can choose $\theta =\pi/2$. Then by introducing  two constants of motion $l$ and $\gamma$, as follows \cite{Boyer}
\begin{eqnarray}\label{44}\notag
p_{\varphi}&=&\frac{\partial \mathcal{L}}{\partial \dot{\varphi}}=r(s)^2 \dot{\varphi}=l   \\\notag
p_{t}&=&\frac{\partial \mathcal{L}}{\partial \dot{t}}=-\left(1-\frac{2G_{0}r^{2}(s)}{r^{3}(s)+\tilde{\omega} G_{0}\left(r(s)+\gamma G_{0}M\right)}\right)\dot{t}\\
&=&-\gamma.
\end{eqnarray}

Furthermore we can introduce a new variable  $r=1/u(\varphi)$, which suggests that
\begin{equation}\label{46}
\frac{\dot{r}}{\dot{\varphi}}=\frac{\mathrm{d}r}{\mathrm{d}\varphi}=-\frac{1}{u^2}\frac{\mathrm{d}u}{\mathrm{d}\varphi}.
\end{equation}

Thus making use of eqs. (32),(34) and (35) and choosing $\gamma=1$ \cite{Boyer} and using the condition that $\varphi$ is measured from the point of closest approach, namely $u(\varphi=0)=u_{max}=1/r_{min}=1/b$ \cite{lorio}, and one can show that for the second constant $l=b$. Then, we get the following differential equation, 
\begin{widetext}
\begin{eqnarray}\notag
-\frac{\Xi(u)^2}{2 u^4 b^2 \zeta(u)^2}+\frac{\Xi(u) G_0 M}{u^6 b^2 \zeta(u)^2}+\frac{\left(\frac{\mathrm{d}u}{\mathrm{d}\varphi}\right)^2 }{2 u^4 \left(1-\frac{2 G_0 M}{u^2 \Xi(u)}\right)}+\frac{1}{2 u^2}=0
\end{eqnarray}
\end{widetext}
where 
\begin{equation}
\Xi(u)=\frac{1}{u^3}+\frac{\tilde{\omega}G_0}{u}+\tilde{\omega}G_0^2 \gamma M
\end{equation}
\begin{equation}
\zeta(u)=\tilde{\omega}G_0^2 \gamma M-\frac{2G_0 M}{u^2}+\frac{1}{u^3}+\frac{\tilde{\omega}G_0}{u}
\end{equation}

We can solve the last equation for $\mathrm{d}u/\mathrm{d}\varphi$, and then to integrate with respect to $u$, to find the deflection angle. By following similar arguments given in Ref. \cite{Boyer}, the deflection angle can be obtained by considering the following integral 
\begin{equation}
\Delta \varphi= 2 \int_0 ^{1/b}  A(u)  \,\mathrm{d}u,
\end{equation}
where $A(u)$ is a complicated function which is obtained by considering Taylor expansion series around $\omega$ and $M$. This function is found to be
\begin{equation}
A(u)=\frac{30 \pi b \left[b^2 u^2(1+G_0 Mu)-1\right]-167 u^5 G_0^2 b^3 M\hbar}{30 \pi (b^2 u^2-1)\sqrt{1-b^2 u^2}},
\end{equation}

Carrying out the integration and keeping only the terms of order $1/b$, and $1/b^{2}$, we can express the final result as 
\begin{equation}
\Delta \varphi =\pi+\hat{\alpha}
\end{equation}
where $\hat{\alpha}$ is the deflection angle.  Thus we find the following result for the deflection angle in the weak deflection approximation to second order terms,  
\begin{equation}
\hat{\alpha} \simeq\frac{4M}{b}-\frac{1336 MG_{0}^{2}\hbar}{45 \,\pi \, b^{3}}.
\end{equation}

\subsection{Quantum corrected black hole in Bohmian quantum mechanics} 
Let us perform similar computations to the case of metric (25). The Lagrangian in this case reads
\begin{widetext}
\begin{equation}\label{43}
\mathcal{L}=-\frac{1}{2}\left(1-\frac{2M)}{r(s)}+\frac{\hbar \eta }{r^2(s)}\right)\dot{t}^2+\frac{\dot{r}^2}{2\left(1-\frac{2M}{r(s)}\right)+\frac{\hbar \eta }{r^2(s)}}+\frac{1}{2}r(s)^2 \left(\dot{\theta}^2+ \sin^{2}\theta \dot{\varphi}^2\right).
\end{equation}
\end{widetext}

In analogous way we define two constants of motion $l$ and $\gamma$ as following \cite{Boyer}
\begin{eqnarray}\label{44}\notag
p_{\varphi}&=&\frac{\partial \mathcal{L}}{\partial \dot{\varphi}}=r(s)^2 \dot{\varphi}=l,\\
p_{t}&=&\frac{\partial \mathcal{L}}{\partial \dot{t}}=-\left(1-\frac{2M}{r(s)}+\frac{\hbar \eta}{r(s)^2}\right)\dot{t}=-\gamma.
\end{eqnarray}

Using the above equations and choosing $\gamma=1$ \cite{Boyer} and using the condition that $\varphi$ is measured from the point of closest approach, namely $u(\varphi=0)=u_{max}=1/r_{min}=1/b$ \cite{lorio}, and one can show that for the second constant
\begin{equation}
l=\frac{b^3+b^2 M-b\eta \hbar -2 \eta \hbar M}{b \sqrt{1- \eta \hbar}},
\end{equation}
then, we get the following differential equation
\begin{widetext}
\begin{eqnarray}\notag
-\frac{M}{u^7 \Xi^2 \zeta^2 }-\frac{\hbar \eta}{2 u^6 \Xi^2 \zeta^2}-\frac{1}{2 u^8 \Xi^2 \zeta^2}+\frac{\left(\frac{\mathrm{d}u}{\mathrm{d}\varphi}\right)^2 }{u^4 \left(2 u^2 \hbar \eta-4Mu+2\right)}+\frac{1}{2 u^2}=0
\end{eqnarray}
\end{widetext}
where 
\begin{equation}
\Xi(u)=\frac{b^3+b^2M-b\eta \hbar-2 \eta \hbar M}{b \sqrt{b^2-\hbar \eta}},
\end{equation}
and
\begin{equation}
\zeta(u)=\hbar \eta -\frac{2M}{u}+\frac{1}{u^2}.
\end{equation}

According to Ref. \cite{Boyer}, the deflection angle can be obtained by considering the following integral 
\begin{equation}
\Delta \varphi= 2 \int_0 ^{1/b}  B(u)  \,\mathrm{d}u,
\end{equation}
where $B(u)$ is a complicated function which is obtained by considering Taylor expansion series around $\hbar$ and $M$. As before we have choosen  $\gamma=1$ and $l=b$. Then, we get the following differential equation, 
\begin{equation}
B(u)=\frac{\Delta b^4+\Theta b^3+\Phi b^2-3 \hbar \eta (Mu+1/3)-3\eta \hbar M}{2 b^2 (bu+1)\sqrt{1-u^2 b^2}}
\end{equation}
\begin{eqnarray}\notag
\Delta&=&-3 \eta \hbar Mu^4-\eta \hbar u^3+2Mu^2+2u,\\\notag
\Theta&=&-3 \eta \hbar Mu^3-\eta \hbar u^2+2Mu+2,\\
\Phi&=&-6 \eta \hbar Mu^2-\eta \hbar u+2M.
\end{eqnarray}

Finally we solve the last integral which can be written as 
\begin{equation}
\Delta \varphi =\pi+\hat{\alpha}
\end{equation}
where $\hat{\alpha}$ is the deflection angle. As before we shall consider only the terms of order $1/b$, and $1/b^{2}$, we can express the final result as 
\begin{equation}
\hat{\alpha} \simeq\frac{4M}{b}-\frac{3 \pi \hbar \eta}{4 \,b^{2}}.
\end{equation}

\bigskip

\section{Conclusion}

In this paper, we have calculated the deflection angle by the renormalization group improved Schwarzschild black hole and a quantum corrected Schwarzschild black hole in the Bohmian quantum mechanics. We have applied the Gauss--Bonnet theorem to the corresponding optical metrics in both cases and recovered the quantum corrected Gaussian optical curvature. We have shown that the deflection angle is affected by the quantum gravity effects in both cases. Finally we confirm GB computations by using the standard geodesic approach up to the second order terms. Morover we have shown that GB method gives exact result up to the second order term in both  solutions.  The importance of this work is mainly of conceptual nature, namely GB method allows to  compute  even quantum corrections to the light deflection, by considering a domain \textit{outside} of the light ray. Hence the role the global topology on gravitational lensing is important in GB method.


\begin{thebibliography}{99}



\bibitem{weak1} P. Schneider, J. Ehlers, and E. E. Falco, Gravitational
Lenses (Springer-Verlag, Berlin, 1992).

\bibitem{weak2} P. Schneider, C. Kochanek, and J. Wambsganss, Gravitational
Lensing: Strong, Weak and Micro (SpringerVerlag,
Berlin, 2006).


\bibitem{strong1} C. Darwin, Proc. R. Soc. London A \textbf{249}, 180 (1959);
263, 39 (1961).

\bibitem{strong2} R. Atkinson, The Astronomical Journal, \textbf{70}, 517 (1965).


\bibitem{strong3} V. Bozza, Gen. Rel. Grav. \textbf{42}, 2269, 2010

\bibitem{virbhadra1} K. S. Virbhadra, and G. F. R. Ellis, Phys. Rev. D \textbf{65}, 103004 (2002)

\bibitem{virbhadra2} K. S. Virbhadra, and C. R. Keeton, Phys. Rev. D \textbf{77}, 124014 (2008).

\bibitem{virbhadra3} K. S. Virbhadra and G. F. R. Ellis, Phys. Rev. D \textbf{62},084003 (2000)

\bibitem{nandi} Kamal Kanti Nandi, Yuan-Zhong Zhang, Alexander V. Zakharov, Phys.Rev. D \textbf{74} (2006) 024020

\bibitem{peter} Peter K.F. Kuhfittig, Eur. Phys. J. C 74, 2818 (2014)

\bibitem{deandrea} J. P. DeAndrea, and K. M. Alexander, Phys. Rev. D \textbf{89}, 123012 (2014).

\bibitem{cos1} E. Hackmann, B. Hartmann, C. Lämmerzahl and P. Sirimachan, Phys. Rev. D 81 (2010) 064016

\bibitem{cos2} E. Hackmann, B. Hartmann, C. Lämmerzahl and P. Sirimachan, Phys. Rev. D 82 (2010) 044024 

\bibitem{sereno1} Sereno, Phys. Rev. D \textbf{69} 023002 (2004)

\bibitem{bozza1} V.Bozza, Phys. Rev. D 67, 103006 (2003)


\bibitem{bozza2} V. Bozza, F. De Luca, G. Scarpetta, and M. Sereno, Phys. Rev. D 72, 083003 (2005)

\bibitem{sereno2} M. Sereno, F. De Luca, Phys.Rev. D \textbf{74} (2006) 123009

\bibitem{pal} Supratik Pal, Sayan Kar, Class. Quant. Grav. 25:045003 (2008)

\bibitem{eiroa1} Ernesto F. Eiroa, Carlos M. Sendra,  Phys.Rev.D, 86:083009, 2012

\bibitem{eiroa2} Ernesto F. Eiroa, Carlos M. Sendra, Eur. Phys. J. C (2014) \textbf{74}, 3171


\bibitem{gibbons1}
Gibbons and Werner, Class. and Quantum Grav. \textbf{25} 235009 (2008)

\bibitem{gibbons2} Gibbons G. W., Herdeiro C. A. R., Warnick C., Werner M. C., Phys. Rev. D \textbf{79} 044022 (2009)

\bibitem{werner}
M.C. Werner, Gen. Rel. Grav. \textbf{44} 3047-3057 (2012)


\bibitem{kimet1} K. Jusufi, Astrophys. Space Sci. \textbf{361}, 24, 2016

\bibitem{kimet2} K. Jusufi, Eur. Phys. J. C (2016) 76:332

\bibitem{kimet3} Kimet Jusufi, Marcus C. Werner, Ayan Banerjee, Ali Ovgon,  Phys. Rev. D \textbf{95}, 104012, 2017

\bibitem{asahi} Asahi Ishihara, Yusuke Suzuki, Toshiaki Ono, Takao Kitamura, Hideki Asada, 	Phys. Rev. D 94, 084015 (2016)

\bibitem{bohr} N. E. J. Bjerrum-Bohr, John F. Donoghue, Barry R. Holstein, Ludovic Planté, Pierre Vanhove, Phys. Rev. Lett. 114, 061301 (2015)

\bibitem{li} Xin Li, Zhe Chang, Phys.Rev. D 82:124009, 2010

\bibitem{sahu} Satyabrata Sahu, Kinjalk Lochan, D. Narasimha, 	Phys. Rev. D 91, 063001 (2015)

\bibitem{sumanta1} Sourav Bhattacharya, Sumanta Chakraborty, arXiv:1607.03693

\bibitem{sumanta2} Sumanta Chakraborty, Soumitra Sengupta, Phys. Rev. D 89, 026003 (2014)
\bibitem{Born-Infeld1} Shao-Wen Wei, Ke Yang, Yu-Xiao Liu, Eur. Phys. J. C (2015) 75:253

\bibitem{Born-Infeld2} Hajime Sotani, Umpei Miyamoto, Phys. Rev. D 92, 044052 (2015)

\bibitem{qcsh0} R. Torres, F. Fayos, O. Lorente-Espin, 	Phys. Lett. B 720 p.198 (2013)

\bibitem{qcsh1} A. Bonanno, M. Reuter, Phys.Rev. D 62(2000)043008.

\bibitem{qcsh2} A. Bonanno, M. Reuter, Phys.Rev. D 73(2006)083005.

\bibitem{qcsh3} N. E. J.Bjerrum--Bohr, J. F. Donoghue, B.R. Holstein, Phys. Rev. D 68(2003)084005

\bibitem{novello} M. Novello, V. A. De Lorenci, J. M. Salim, R. Klippert, Phys. Rev. D 61 (2000) 045001

\bibitem{bm1} S. Das, Phys. Rev. D, \textbf{89}, 084068, 2014.

\bibitem{bm2} Ahmed Farag Ali, Saurya Das, Phys. Lett. B \textbf{741} (2015) 276

\bibitem{bm3} A.F. Ali, M.M. Khalil, Nucl. Phys. B, \textbf{909},
173-185 (2016).

\bibitem{odintsov} S. Falkenberg,  S. D. Odintsov, Int. J. Mod. Phys. A, 13, 607-623, 1998


\bibitem{Boyer} R. H. Boyer, R. W. Lindquist, J. Math. Phys. \textbf{8}, 265 (1967).


\bibitem{lorio} L. Iorio, Nuovo Cim. B, \textbf{118}, 249 (2003).

\end{thebibliography}
\end{document}